\documentclass[prl,aps,twocolumn,showpacs,floatfix]{revtex4-1}
\usepackage{tabularx}
\usepackage{amsmath}
\usepackage{amssymb}
\usepackage{hyperref}
\usepackage{graphicx}
\usepackage{dcolumn}
\usepackage{bm}

\def\be{\begin{equation}}
\def\ee{\end{equation}}
\def\bea{\begin{eqnarray}}
\def\eea{\end{eqnarray}}
\def\bse{\begin{subequations}}
\def\ese{\end{subequations}}

\renewcommand{\v}[1]{{\bf #1}}

\def\be{\begin{eqnarray}}
\def\ee{\end{eqnarray}}

\begin{document}

\title{Observation of Zitterbewegung in a spin-orbit coupled Bose-Einstein
condensate}
\author{Chunlei Qu$^{1}$}
\thanks{These authors contributed equally to this work}
\author{Chris Hamner$^{2}$}
\thanks{These authors contributed equally to this work}
\author{Ming Gong$^{1}$}
\author{Chuanwei Zhang$^{1}$}
\thanks{Email: chuanwei.zhang@utdallas.edu}
\author{Peter Engels$^{2}$}
\thanks{Email: engels@wsu.edu}

\begin{abstract}
Spin-orbit coupled ultra-cold atoms provide an intriguing new avenue for the
study of rich spin dynamics in superfluids. In this Letter, we observe
Zitterbewegung, the simultaneous velocity (thus position) and spin
oscillations, of neutral atoms between two spin-orbit coupled bands in a
Bose-Einstein condensate (BEC) through sudden quantum quenches of the
Hamiltonian. The observed Zitterbewegung oscillations are perfect on a short
time scale but gradually damp out on a long time scale, followed by sudden
and strong heating of the BEC. As an application, we also demonstrate how
Zitterbewegung oscillations can be exploited to populate the upper
spin-orbit band, and observe a subsequent dipole motion. Our experimental
results are corroborated by a theoretical and numerical analysis and
showcase the great flexibility that ultra-cold atoms provide for
investigating rich spin dynamics in superfluids.
\end{abstract}

\affiliation{$^{1}$Department of Physics, the University of Texas at Dallas, Richardson,
TX 75080 USA \\
$^{2}$Department of Physics and Astronomy, Washington State University,
Pullman, Washington 99164, USA}
\maketitle

\emph{Introduction}.---The Zitterbewegung (ZB) oscillation, first predicted
by Schr\"{o}dinger in 1930 \cite{ZB1930} for relativistic Dirac electrons,
describes the fast oscillation or trembling motion of electrons arising from
the interference between particle and hole components of Dirac spinors.
Although fundamentally important, the ZB oscillation is difficult to observe
in real particles. In the past eight decades, analogs of the ZB oscillation
have been predicted to exist in various physical systems \cite%
{ZBgraphene,ZBGeim,ZBTI,ZBsemiconductor,ZBatom,ZBdamp,Ions-Solano}, ranging
from solid state (\textit{e.g.}, semiconductor quantum wells) to trapped
cold atoms, but experimentally a ZB analog has only recently been observed
using trapped ions as a quantum emulator of the Dirac equation \cite{ZBexp}.
A crucial ingredient for the ZB oscillation is the coupling between spin and
linear momentum of particles, leading to simultaneous velocity and position
oscillations accompanying the spin oscillation, which distinguishes ZB from
Rabi oscillations where spin oscillations between two bands do not induce
velocity and position oscillations.

Ultra-cold atomic gases provide a very promising setting for emulating
interesting quantum phenomena because of the high tunability of system
parameters as well as the direct imaging of atomic velocities and positions.
For instance, recent experiments have succeeded in the realization of
spin-orbit (SO) coupling in Bose-Einstein condensates (BECs) and degenerate
Fermi gases (DFGs) \cite{Lin1,Pan - Collective Dipole modes,Zwierlein - Spin
Injection,Jing Zhang DFG SO,Dalibard}. While SO coupling plays a prominent
role in many important condensed-matter phenomena \cite{SOCM1,SOCM2,SOCM3},
its realization in neutral atom superfluids is novel and provides a powerful
experimental platform due to a rich ground state phase diagram, intriguing
equilibrium and non-equilibrium spin dynamics and the presence of many-body
interactions \cite%
{gs1,gs2,gs3,gs4,gs5,gs6,gs7,gs8,gs9,gs10,gs11,gs12,gs13,gs14}.

As we show in this Letter SO coupling in a BEC makes it possible to observe
ZB oscillations in neutral atomic gases. To induce ZB oscillation in a SO
coupled BEC, we exploit quantum quenches of the Hamiltonian. The study of
quenches and many-body dynamics far from equilibrium has emerged as an
important frontier in many branches of physics \cite%
{quench2,quench3,quench,Ising,Ising2,Ising3,Fermion,BH,Iso}, including cold
atomic gases. We observe short-time coherent ZB oscillations as well as long
time ZB damping as a consequence of such quenches. Our main observations are
the following:

(I) On a short time scale ($\sim 1$ ms), a quantum quench couples two SO
bands, leading to simultaneous spin and velocity oscillations of the BEC
that can be interpreted as ZB oscillation. Here the two SO bands effectively
mimic the particle and hole branches of the Dirac equation, and the
oscillation frequency is determined by the energy splitting between the two
bands.

(II) On a long time scale ($\sim 10$ ms), the amplitude of the ZB
oscillation damps out because of the diminishing overlap of the two
wavepackets as they move with different group velocities in the two bands.
The many-body interactions between atoms reduce the damping of the
oscillation amplitude. After the ZB oscillation damps out, a subsequent
dipole motion is accompanied by sudden and strong heating of the BEC.

(III) The ZB oscillation can be used to load a BEC into the upper SO band.
As a result, we observe dipole motion of the BEC in the upper band as well
as the accompanying change in spin composition.

\begin{figure}[tbp]
\centering
\includegraphics[width=3.2in]{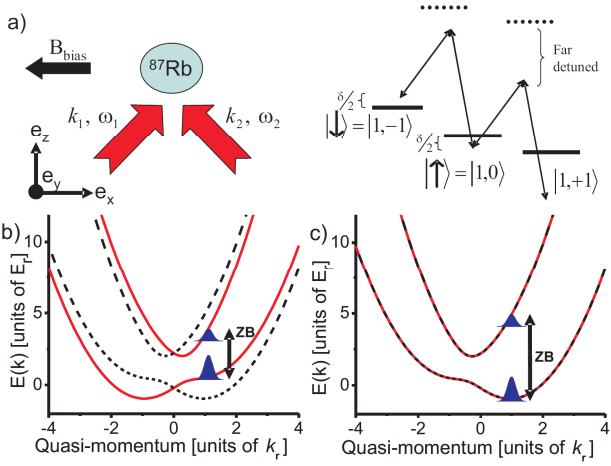}
\caption{(Color Online). (a) Experimental configuration for the creation of
spin-orbit coupling in the $F=1$ manifold of a $^{87}$Rb BEC. (b) Typical
band structure before (dashed-black) and after (solid-red) quenching the
system by jumping the detuning $\protect\delta $. The wavepackets
symbolically show the wavefunction directly after the quench. The short time
dynamics are dominated by ZB oscillations. (c) Similar to (b) but for a jump
of the relative phase between the two Raman beams. The band structure is
unaltered by the phase jump. }
\label{intro}
\end{figure}

\emph{Experimental Methods and theoretical model}.---Our experiments are
conducted with BECs of $^{87}$Rb of about $1-2\times 10^{5}$ atoms confined
in a trapping potential with trapping frequency $\omega _{x,y,z}=2\pi \times
\{20-40,174,120\}$~Hz, where the value of $\omega _{x}$ depends on the
intensity of the Raman beams (thus $\Omega $) as well as on a crossed dipole
beam. Two crossed Raman lasers with wavelengths near $\lambda =784$ nm
propagate along the $\mathbf{e}_{x}\pm \mathbf{e}_{y}$ direction (relative
angle $=90^{\circ }$), respectively. We apply a magnetic bias field of 10~G
in the $\mathbf{e}_{x}$ direction (SO coupling direction) as shown in Fig.~%
\ref{intro}a. The resulting quadratic Zeeman splitting for the $F=1$
manifold $\epsilon _{z}$ is $7.6E_{r}$, which is sufficiently large such
that the contribution of the hyperfine state $|1,1\rangle $ can mostly be
neglected, yielding an effective spin-1/2 system with the pseudo-spins
defined as $|\uparrow \rangle \equiv |1,0\rangle $ and $|\downarrow \rangle
\equiv |1,-1\rangle $. Here $E_{r}=\hbar ^{2}k_{r}^{2}/2m=\hbar \times 2\pi
\times 1.866$~kHz is the recoil energy and $\hbar k_{r}=\sqrt{2}\pi \hbar
/\lambda $ is the recoil momentum.

In the pseudo-spin basis, the dynamics of the BEC can be theoretically
described by an effective two-band Gross-Pitaevskii (G-P) equation with the
corresponding Hamiltonian $H=H_{0}+V_{t}+H_{I}$. Here the single atom
Hamiltonian is given by\textbf{\ }

\begin{equation}
H_{0}=%
\begin{pmatrix}
\frac{\hbar ^{2}}{2m}(\mathbf{k}+k_{r}\mathbf{e}_{x})^{2}+{\frac{\delta }{2}}
& {\frac{\Omega }{2}} \\
{\frac{\Omega ^{\ast }}{2}} & \frac{\hbar ^{2}}{2m}(\mathbf{k}-k_{r}\mathbf{e%
}_{x})^{2}-{\frac{\delta }{2}}%
\end{pmatrix}
\label{eq-Heff}
\end{equation}%
after a local pseudo-spin rotation \cite{Lin1}. For small $k$, $H_{0}$
reduces to the Dirac equation \cite{ZBexp}. $\Omega $ is the Raman coupling
strength, and $\delta $ is the detuning of the Raman transition, effectively
acting like a Zeeman field. The experimental results are accompanied by the
G-P numerical simulation performed in a 2D cigar shaped geometry, in which $%
\mathbf{k}=k_{x}\mathbf{e}_{x}+k_{y}\mathbf{e}_{y}$ is the quasi-momentum of
atoms. The harmonic trapping potential is $V_{t}=m\omega
_{x}^{2}x^{2}/2+m\omega _{y}^{2}y^{2}/2${. }The many-body interactions
between atoms are described by the nonlinear term
\begin{equation}
H_{I}=\text{diag}\left( \sum\nolimits_{\sigma =\uparrow ,\downarrow
}g_{\uparrow \sigma }|\psi _{\sigma }|^{2},\sum\nolimits_{\sigma =\uparrow
,\downarrow }g_{\downarrow \sigma }|\psi _{\sigma }|^{2}\right) {,}
\label{2DGP}
\end{equation}%
where the effective 2D interaction parameters are given by $g_{\uparrow
\uparrow }={\frac{2\sqrt{2\pi }\hbar ^{2}Nc_{0}}{ma_{z}}}$ and $g_{\uparrow
\downarrow }=g_{\downarrow \uparrow }=g_{\downarrow \downarrow }={\frac{2%
\sqrt{2\pi }\hbar ^{2}N(c_{0}+c_{2})}{ma_{z}}}$, with the spin-dependent 3D
\textit{s}-wave scattering lengths $c_{0}$ and $c_{0}+c_{2}$ for Rb atoms ($%
c_{2}=-0.46a_{0}$ and $c_{0}=100.86a_{0}$). $a_{0}$ is Bohr radius, and $%
a_{z}=\sqrt{\frac{\hbar }{m\omega _{z}}}$ is the harmonic oscillator length.

In our quench experiments, we first prepare the system in the ground state
with a finite detuning $\delta $ and an initial quasi-momentum $k_{x}$ near $%
k_{r}$. The system is quenched by either a sudden jump of the Zeeman field
from $\delta $ to $-\delta $ (via a frequency jump of the lasers), or by a
sudden phase jump of $\pi $ of the Raman field (which is equivalent to
jumping $\Omega $ to $-\Omega $ in the Hamiltonian (\ref{eq-Heff})) (Figs.~%
\ref{intro}b,c). The jumps in $\delta $ or the sign of $\Omega $ are
effected in less than 10~$\mu $s which is much shorter than the ZB
oscillation period and any relevant system dynamics timescale. After the
quench, we allow the system to evolve for a given evolution time before
starting the imaging procedure. The imaging procedure consists of jumping
off the Raman coupling and external confinement, allowing $11.5~$ms
time-of-flight in the presence of a Stern-Gerlach field that separates the
bare states, and imaging the bare spin states along the $\mathbf{e}_{z}$
direction. The images are oriented such that the horizontal axis coincides
with the direction of the momentum transfer of the Raman coupling.

\begin{figure}[tbp]
\centering
\includegraphics[width=3.2in]{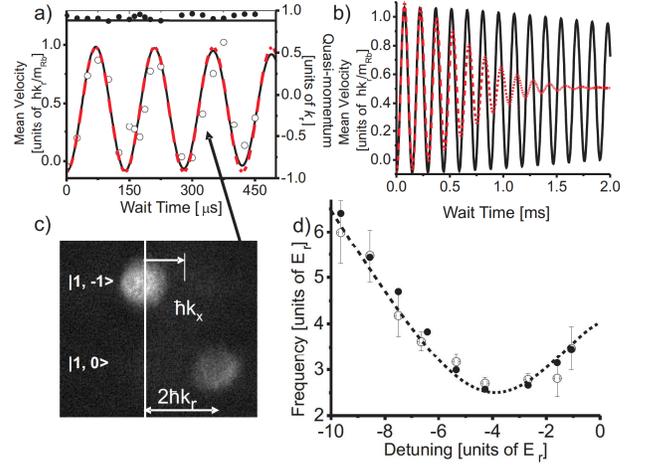}
\caption{(color online)(a) Experimental observation of the ZB oscillation of
$\left\langle v_{x}\right\rangle $ for $\Omega =2.5E_{r}$ and $\protect%
\delta $ jumped from $6.42E_{r}$ to $-6.42E_{r}$. Experimental data (open
circles) is shown overlaid with results of numerical simulations of G-P
equation (black line) and analytic prediction based on effective two band
Hamiltonian (dashed red line). Experimental (black dot) and numerical (solid
line) quasi-momenta are shown in top part of plot. (b) $\left\langle
v_{x}\right\rangle $ for numerical simulation of experimental parameters
with (solid line, same as (a)) and without (dashed line) interactions. (c)
Experimental image taken at $t=325{\protect\mu }s$ showing the Stern-Gerlach
separation and 2$\hbar $k$_{r}$ photon momentum separation of the bare
states. (d) Experimental (open circles) and numerical results (filled
circles) for ZB oscillation frequency vs. $\protect\delta $ after quench.
The dashed line shows the band splitting calculated from the effective two
band model for $\Omega =2.5E_{r}$. The experimental error bars are
determined from fit uncertainties, shot to shot variations of $k_{x}$, and
calibration uncertainty of $\Omega $. }
\label{ZB}
\end{figure}

\emph{Zitterbewegung oscillation}.---In order to exhibit a velocity
oscillation in ZB, a spin oscillation between two SO bands is needed, which
is realized by the quantum quenches in our experiments. After the quench,
the initial wave function is no longer an eigenstate of the new Hamiltonian,
therefore there is a fraction of BEC projected into the upper band. The
components in the upper and lower bands beat against each other, leading to
spin oscillations in the bare state basis. From the single particle
Hamiltonian (\ref{eq-Heff}), the center of mass motion of the atomic
wavepacket is given by \cite{SI}
\begin{equation}
\left\langle v_{x}\right\rangle =\hbar (q+k_{r}\left\langle \sigma
_{z}\right\rangle )/m=\left( N_{\uparrow }v_{\uparrow }+N_{\downarrow
}v_{\downarrow }\right) /N,  \label{velocity}
\end{equation}%
where $v_{\uparrow }=\hbar (q+k_{r})/m$ and $v_{\downarrow }=\hbar
(q-k_{r})/m$ are the velocities of the two components. Therefore the
oscillating spin polarization $\langle \sigma _{z}\rangle $ leads to the
oscillation of the velocity $\left\langle v_{x}\right\rangle $ (the
quasi-momentum $q=\left\langle k_{x}\right\rangle $ is roughly a constant on
the short time scale). The velocity is directly observed in our experiment
shown in Fig.~\ref{ZB}a following a jump of $\delta $ from $6.42E_{r}$ to $%
-6.42E_{r}$ with $\Omega =2.5E_{r}$. Here the dynamics are characterized by
a rapid population oscillation between the $|\downarrow >$ and $|\uparrow >$
spin states with a momentum transfer of 2$\hbar k_{r}$ as seen in Fig.~\ref%
{ZB}c.

The frequency of the ZB oscillation is determined by the energy splitting
between two SO bands. For the chosen parameters the oscillations occur at a
frequency of 3.62$E_{r}$ and can be observed for many cycles. The
quasi-momentum remains relatively constant over this time scale. The
dependence of the velocity oscillation frequency on the parameter $\delta $
is plotted in Fig.~\ref{ZB}d along with the band excitation frequency and
the results from numerical simulations of the nonlinear G-P equation \cite%
{SI}. Clearly the oscillation frequency is well described by the energy
splitting between the lower and upper band. The velocity oscillation
amplitude is bounded by $\sim \hbar k_{r}/m\approx 4.14$ mm/s \cite{SI,
AmplitudeFootNote}. Due to the similarity between the effective two-band
model in Eq. (\ref{eq-Heff}) and the Dirac-like equation, the observed
velocity oscillation is a low-temperature analog to the well-known ZB
oscillation theoretically studied in various systems \cite{ZBgraphene,
ZBGeim, ZBTI, ZBsemiconductor, ZBatom, ZBdamp, Ions-Solano}, but only
observed previously in \cite{ZBexp}. The occurrence of such oscillations is
not unique to quenches of $\delta $. For instance, we have observed similar
velocity oscillation for jumping $\Omega $ to $-\Omega $ (see Fig. \ref%
{fig-upper band}a). Note that without SO coupling, Eq. (\ref{velocity})
becomes $\left\langle v_{x}\right\rangle =\hbar q$, and is a constant. In
this case the velocity (and thus the position) is independent of the spin,
and the spin oscillation does not induce the velocity and position
oscillations of atoms, leading to a standard Rabi oscillation, instead of a
ZB oscillation.

An ideal sinusoidal ZB oscillation is possible only when a single momentum
is involved in the initial ground state. However, in a realistic system in a
harmonic trap, the single particle ground state has a small spread of the
momentum around the minimum of the band which leads to damping on the
timescale of a few oscillations \cite{ZBatom}. Furthermore, the ZB
oscillation is only present when the wave functions in the two SO bands have
significant overlap in real space. When the wave packets in the two SO bands
move with different group velocities, they start to separate in real space,
leading to strong damping \cite{ZBdamp} on a longer timescale. Many-body
effects also affect the damping by expanding the wave function in real space
and narrowing it in momentum space, leading to a reduced damping effect on
both short and long time scales as seen in Figs.~\ref{ZB}b and \ref%
{fig-delayed onset}b.

\begin{figure}[tbp]
\centering
\includegraphics[width=3.2in]{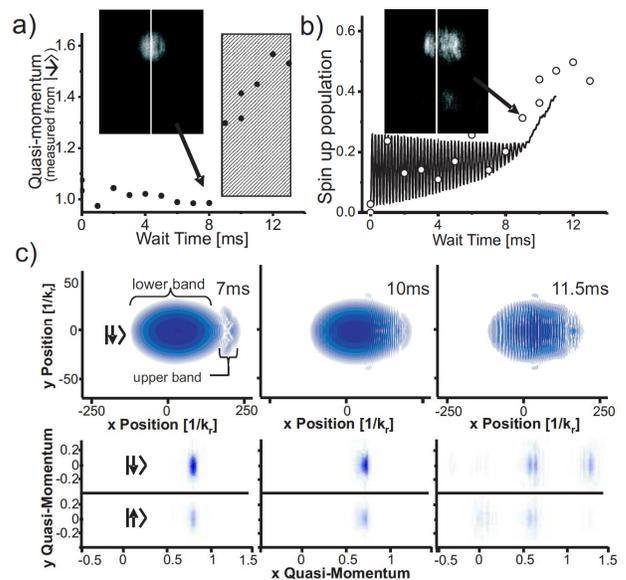}
\caption{(Color Online). BEC loaded to $\Omega =1.8E_{r}$ and $\protect%
\delta =1.6E_{r}$, followed by jump of $\protect\delta $ to $-1.6E_{r}$. a)
Experimentally observed quasi-momentum for the component of the BEC near $%
k_{x}=k_{r}$. Grey region indicates onset of heating. b) Corresponding
experimental (open circles) and numerical results (solid line) for spin
composition of the BEC. Insets of a) and b) show images of the BEC at $t=8ms$
and $t=9ms$ (i.e. just before and after onset of heating). Vertical line
indicates zero kinetic momentum. c) Numerical simulations show the real and
momentum space composition for $7ms$ , $10ms$, $11.5ms$, respectively.}
\label{fig-delayed onset}
\end{figure}

An interesting observation is the sudden increase in motion, after $\sim 8$
ms, and the subsequent rapid heating of the BEC, see Figs.~\ref{fig-delayed
onset}a and \ref{fig-delayed onset}b. Note that the ZB oscillation, which
typically damps out on the time scale of a few ms, is present in these cases
as well but is not resolved in the experimental data shown in this figure
due to the chosen 1~ms time steps. The delayed onset of the dipole motion is
also observed in the numerics and is related to the geometry of the band
structure. For a small value of the Raman coupling, the region of the lower
band after the quench of $\delta $ near the initial BEC quasimomentum is
relatively flat, implying a small initial group velocity. We have verified
in our numerics that for a larger Raman coupling, there is no such delayed
onset. The observed strong heating is also related to the geometry of the
band structure: Unlike the BEC in the lower band, the BEC in the upper band
has a large initial group velocity. By $7$ ms the ZB oscillation damps out
after there is no longer any overlap of the components in the two bands in
real space. Subsequently the wave packet in the upper band turns around
because of the trap, and its collision with atoms in the lower band leads to
excitations of many momentum modes \cite{instability}, as seen in Fig.~\ref%
{fig-delayed onset}c for 11.5 ms. The fact that the BEC in the lower band
enters a negative effective mass region of the band structure may also
contribute to the excitations.

After the ZB oscillations damp out, the BEC continues to move within the
quenched band structure and performs dipole oscillations \cite{Lin Synthetic
electric field,Pan - Collective Dipole modes} with strong spin relaxation
\cite{SI}. Such subsequent spin relaxation to the potential minimum in the
presence of many-body interaction between atoms is different from that in
solid state systems originating from the scattering from impurities. In our
experiments, the dipole oscillations occur as a consequence of the quantum
quench, with the initial wave packets in both lower and upper SO bands. The
stability and relaxation of this large amplitude dipole oscillation has not
been studied before. An example of the dynamics in the subsequent dipole
oscillations is shown in the supplementary material \cite{SI}.

\begin{figure}[tbp]
\centering
\includegraphics[width=3.2in]{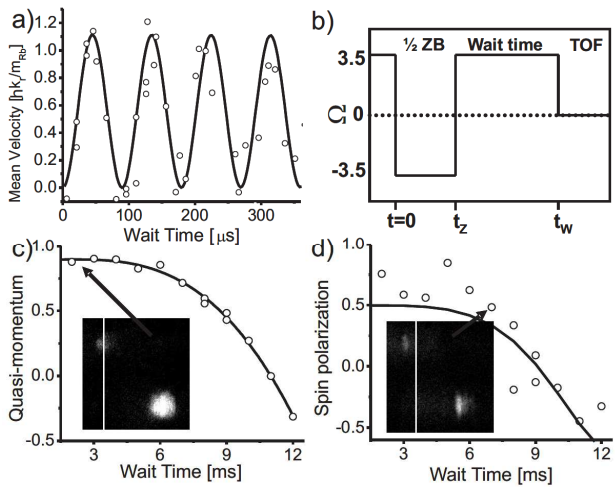}
\caption{Loading of the upper band via two phase jumps separated by $40{%
\protect\mu }s$ where $\Omega =3.5E_{r}$ and $\protect\delta =1.6E_{r}$.
a)Experimentally observed ZB oscillation (open circles) after a single phase
jump, and corresponding numerical simulations (solid line) for experimental
conditions. b) The time sequence. c-d) Experimentally observed
quasi-momentum and spin polarization, respectively. Insets are experimental
images taken during the evolution where the vertical line indicates zero
kinetic momentum.}
\label{fig-upper band}
\end{figure}

\emph{Loading a BEC into the upper band using ZB}.---In our discussion so
far, we have concentrated on the ZB oscillation following a single quantum
quench. In the following we demonstrate how a sequence of quenches and the
resulting ZB oscillations can be exploited to load the BEC nearly entirely
into the upper band and thus forms a pathway to studying upper band
dynamics. We demonstrate this by two phase jumps of the Raman lasers. The
first jump from $\Omega $ to $-\Omega $ starts ZB oscillation, after which
we wait for half a cycle until the vast majority of the population has been
transferred from $|\downarrow \rangle $ to $|\uparrow \rangle $. This is
followed by a second jump from $-\Omega $ to $\Omega $ (Fig.~\ref{fig-upper
band}b). We experimentally investigate this procedure using $\Omega
=3.5E_{r} $ and $\delta $ $=1.6E_{r}$, inciting large ZB oscillation
amplitudes by a phase jump of $\pi $ of the Raman fields, as seen in Fig.~%
\ref{fig-upper band}a. When the phase change is reversed by a second jump at
$t_{Z}=40~\mu $s, we load approximately 80$\%$ of the BEC into the upper
band near $k_{x}=k_{r}$. This is in agreement with an argument based on the
corresponding transformations in the Bloch sphere. As the population
transfer is not unity for these chosen parameters, there will be a small
residual ZB oscillation after the second jump. Allowing an evolution time $%
t_{W}-t_{Z}$, the BEC undergoes dipole motion while it gradually melts. The
quasi-momentum and the spin polarization, defined as $(N_{\uparrow
}-N_{\downarrow })/(N_{\uparrow }+N_{\downarrow })$ for the component of the
BEC in the upper band are plotted in Figs.~\ref{fig-upper band}c and \ref%
{fig-upper band}d. The change in spin polarization with quasi-momentum is
consistent with the prediction from the single particle model. For this
calculation we fit the quasi-momentum measurement with a polynomial (shown
as a line in Fig. \ref{fig-upper band}c). Using the single particle band
structure we then calculate the corresponding spin polarization (shown as a
line in Fig.~\ref{fig-upper band}d). The good agreement with the
experimental data indicates the spinfulness of the upper band.

\emph{Summary}.---In summary, we observe ZB oscillations of neutral atoms
for the first time through quenching a spin-orbit coupled BEC. We find that
many-body interaction between atoms plays an important role for ZB
oscillations and their decay. The results presented in this work showcase
the exceptional flexibility that cold atoms provide for the study of quantum
spin dynamics in spin-orbit coupled superfluids. The rich physics accessible
by rapid quenches of various system parameters offer exciting outlooks for
further studies, such as upper band dynamics, spin decoherence, etc.

Note: After the completion of this manuscript, a related measurement of ZB
oscillation was posted \cite{SpielmanZB}.

\textit{Acknowledgement:} C.Q., M.G. and C.Z. are supported by ARO
(W911NF-12-1-0334), DARPA-YFA (N66001-10-1-4025), AFOSR (FA9550-11-1-0313),
and NSF-PHY (1104546). C.H. and P.E. acknowledge funding from NSF and ARO.

\begin{appendix}
\begin{widetext}

\section{ZB oscillation after a quench in the Heisenberg picture}

For simplicity, in the following discussion we use the dimensionless
single particle Hamiltonian, \textit{i.e.}, we measure energy in
unit of $E_{r}$, and $k_{x}$ in unit of $k_{r}$. We assume that the
Raman coupling strength is real and a quench of the detuning from
$\delta$ to $-\delta$ is performed.

The Hamiltonian after the quench can be written as
\begin{equation}
H=\left( k_{x}^{2}+1\right) I_{2}+(2k_{x}-\frac{\delta }{2})\sigma _{z}+%
\frac{\Omega }{2}\sigma _{x},
\end{equation}%
where $I_{2}$ is a $2\times 2$ unit matrix. In the Heisenberg
picture, the wave function does not change while the spin operator
evolves as
\begin{eqnarray}
\frac{d\sigma _{z}}{dt} &=&-i[\sigma _{z},H] \\
&=&-i(\sigma _{z}H-H\sigma _{z}) \\
&=&-i(\{\sigma _{z},H\}-2H\sigma _{z}).
\end{eqnarray}%
Using the fact that $\{\sigma _{z},1\}=2\sigma _{z}$, $\{\sigma _{z},\sigma
_{x}\}=0$, $\{\sigma _{z},\sigma _{z}\}=2$, we obtain
\begin{equation*}
\frac{d\sigma _{z}}{dt}=2i\{(H-k_{x}^{2}-1)\sigma _{z}-(2k_{x}-\frac{\delta
}{2})\}.
\end{equation*}%
Integrating this differential equation, we obtain the formal solution for $%
\sigma _{z}$,
\begin{equation}
\sigma _{z}(t)=\frac{2k_{x}-\frac{\delta }{2}}{H-k_{x}^{2}-1}+\left[ \sigma
_{z}(0)-\frac{2k_{x}-\frac{\delta }{2}}{H-k_{x}^{2}-1}\right] \exp \left(
2i(H-k_{x}^{2}-1)t\right) ,
\end{equation}%
where $\sigma _{z}(0)$ is the spin operator at $t=0$. The last term
is a fast oscillating function of time, leading to a fast
oscillating velocity, according to Eq. (3) in the main manuscript.
Such a velocity leads to the oscillation of the position, which is
the Zitterbewegung. \newline

The initial wave function is the ground state of the Hamiltonian before the
quench, and $k_{x} $ is determined by the global minimum of the lower band
structure. The ground state $|\psi (t=0)\rangle =|g\rangle =(\cos {\theta }%
~\sin {\theta })^{T}$, is a function of the system parameters $\delta $, $%
\Omega $ and $k_{x}$. After the quench, the eigenvalues are $E_{\pm
}=k_{x}^{2}+1\pm \frac{\Delta E}{2}$ with $\Delta
E=\sqrt{(4k_{x}-\delta )^{2}+\Omega ^{2}}$, and the corresponding
orthogonal eigenvectors are denoted as $|+\rangle =(\cos {\alpha
}~\sin {\alpha })^{T}$ and $|-\rangle =(\sin {\alpha }~-\cos {\alpha
})^{T}$. Therefore we have $\langle +|g\rangle =\langle g|+\rangle
=\cos (\alpha -\theta )$, $\langle -|g\rangle =\langle g|-\rangle
=\sin (\alpha -\theta )$.

Using the fact that $\langle +|\sigma _{z}(0)|+\rangle =-\langle -|\sigma
_{z}(0)|-\rangle =\cos (2\alpha )$, and $\langle +|\sigma _{z}(0)|-\rangle
=\langle -|\sigma _{z}(0)|+\rangle =\sin (2\alpha )$, we can calculate the
expectation value of the spin polarization operator
\begin{eqnarray}
P &=&\langle \sigma _{z}\rangle =\langle g|\sigma _{z}(t)|g\rangle \\
&=&\langle g|(|+\rangle \langle +|+|-\rangle \langle -|)\sigma
_{z}(t)(|+\rangle \langle +|+|-\rangle \langle -|)|g\rangle \\
&=&\cos ^{2}(\alpha -\theta )\langle +|\sigma _{z}(t)|+\rangle +\sin
^{2}(\alpha -\theta )\langle -|\sigma _{z}(t)|-\rangle \\
&+&\frac{1}{2}\sin (2\alpha -2\theta )\langle -|\sigma _{z}(t)|+\rangle +%
\frac{1}{2}\sin (2\alpha -2\theta )\langle +|\sigma _{z}(t)|-\rangle \\
&=&\cos ^{2}(\alpha -\theta )\left[ \frac{2k_{x}-\frac{\delta }{2}}{%
E_{+}-k_{x}^{2}-1}+\left( \cos (2\alpha )-\frac{2k_{x}-\frac{\delta }{2}}{%
E_{+}-k_{x}^{2}-1}\right) e^{2i(E_{+}-k_{x}^{2}-1)t}\right] \\
&+&\sin ^{2}(\alpha -\theta )\left[ \frac{2k_{x}-\frac{\delta }{2}}{%
E_{-}-k_{x}^{2}-1}+\left( -\cos (2\alpha )-\frac{2k_{x}-\frac{\delta }{2}}{%
E_{-}-k_{x}^{2}-1}\right) e^{2i(E_{-}-k_{x}^{2}-1)t}\right] \\
&+&\frac{1}{2}\sin (2\alpha -2\theta )\sin (2\alpha
)e^{2i(E_{+}-k_{x}^{2}-1)t}+\frac{1}{2}\sin (2\alpha -2\theta )\sin (2\alpha
)e^{2i(E_{-}-k_{x}^{2}-1)t}.
\end{eqnarray}%
If we express the eigenvectors $|+\rangle $ and $|-\rangle $ explicitly
using the system parameters, we find that
\begin{equation}
\cos (2\alpha )=\frac{2k_{x}-\frac{\delta }{2}}{E_{+}-k_{x}^{2}-1}=-\frac{%
2k_{x}-\frac{\delta }{2}}{E_{-}-k_{x}^{2}-1},
\end{equation}%
and the spin polarization
\begin{equation}
P=\frac{N_{\uparrow }-N_{\downarrow }}{N_{\uparrow }+N_{\downarrow
}}=\cos (2\alpha -2\theta )\cos (2\alpha )+\sin (2\alpha -2\theta
)\sin (2\alpha )\cos (\Delta Et).
\end{equation}%
%
%
%.

The ZB oscillation frequency for quenching $\delta $ is (including
units)
\begin{equation}
\hbar \omega _{ZB}=\Delta E=\sqrt{(4k_{x}E_{r}/k_{r}-\delta )^{2}+\Omega ^{2}%
}.
\end{equation}%
and the velocity oscillation term is
\begin{equation*}
\frac{\hbar k_{r}}{m}\sin (2\alpha -2\theta )\sin (2\alpha )\cos (\frac{%
\Delta E}{\hbar }t).
\end{equation*}%
Therefore the maximum amplitude of the position oscillation is
smaller than
\begin{eqnarray}
\frac{\pi }{\omega _{ZB}}\frac{\hbar k_{r}}{m}\sin (2\alpha -2\theta )\sin
(2\alpha ) &\sim &\frac{\pi \hbar }{\Delta E}\frac{\hbar k_{r}}{m} \\
&\sim &\frac{\pi \hbar }{E_{r}}\frac{\hbar k_{r}}{m} \\
&\sim &1/k_{r},
\end{eqnarray}%
which is much smaller than $1$ $\mu $m and is below the imaging
resolution of our experiment.

\section{Long time dipole oscillation}

An example of the dipole oscillation generated by jump of $\delta $ is shown
in Figs. S\ref{fig-dipole oscillations}a and S\ref{fig-dipole oscillations}b.
Here the BEC is loaded with $\Omega =1.4E_{r}$ and $\delta =1.6E_{r}$ and
quenched to $-1.6E_{r}$. After the BEC melts, the cloud re-condenses near
the new global minimum in the presence of continued evaporation (grey region
in Fig. S\ref{fig-dipole oscillations}). A small amplitude dipole oscillation
remains as seen in the plot of the quasi-momentum in Fig. S\ref{fig-dipole
oscillations}a. The spin relaxes to the almost fully spin-polarized state
(Fig. S\ref{fig-dipole oscillations}b).

\begin{figure}[tbp]
\centering
\includegraphics[width=3.2in]{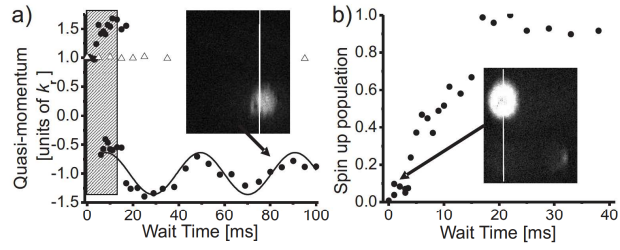}
\caption{Long time dynamics after a quantum quench. The BEC is loaded to $%
\protect\delta =1.6E_{r}$ and $\Omega =1.4E_{r}$, followed by a jump of $%
\protect\delta $ to $-1.6E_{r}$. a) Quasi-momentum of the BEC after
the quench (solid circles), where the gray region indicates strong
heating and rethermalization. For comparison, the quasi-momentum of
the BEC without a quench is also shown (open triangles). The solid
line indicates a fit to the dipole oscillation. b) Spin up
population. The insets are experimental images taken during the
evolution, and the vertical line indicates zero kinetic momentum. }
\label{fig-dipole oscillations}
\end{figure}

\end{widetext}
\end{appendix}

\begin{thebibliography}{99}
\bibitem{AmplitudeFootNote} The velocity oscillation naturally yield the ZB
trembling motion of atoms. However, the amplitude of the oscillation in
position space for our chosen parameters is small (below 1 $\mu $m) and thus
is not directly observable in our experiments.
\end{thebibliography}
\end{document}